\documentclass[%
twocolumn,
showpacs,preprintnumbers,
 amsmath,amssymb,
 aps,
10pt]{revtex4-1}

\usepackage{graphicx,color,ulem}
\usepackage{braket}
\usepackage{hyperref}

\definecolor{battleshipgrey}{rgb}{0.52, 0.52, 0.51}
\definecolor{cadet}{rgb}{0.33, 0.41, 0.47}
\definecolor{charcoal}{rgb}{0.21, 0.27, 0.31}

\hypersetup{
	colorlinks   = true, 
	urlcolor     = battleshipgrey, 
	linkcolor    = cadet, 
	citecolor   = charcoal 
}

\newcommand{\pb}{\mathbf{p}}
\newcommand{\dd}{\mathrm{d}}
\newcommand{\rb}{\mathbf{r}}
\newcommand{\Ab}{\mathbf{A}}

\renewcommand\Re{\mathrm{Re}}
\renewcommand\Im{\mathrm{Im}}

\begin{document}

\preprint{APS/123-QED}

\title{Treating Branch Cuts in 
	Quantum Trajectory Models for Photoelectron Holography}

\author{A. S. Maxwell$^1$, S. V. Popruzhenko$^{2,3}$ and C. Figueira de Morisson Faria$^1$}
\affiliation{$^1$Department of Physics \& Astronomy, University College London Gower Street London  WC1E 6BT, United Kingdom\\$^2$Max Planck Institute for the Physics of Complex Systems, Dresden D-01187, Germany\\$^3$Department of Physics, Voronezh State University, Voronezh 394018, Russia}

\date{\today}

\begin{abstract}

Most implementations of Coulomb-distorted strong-field approaches that contain features such as tunneling and quantum interference use real trajectories in continuum propagation, while a fully consistent approach must use complex trajectories throughout. A key difficulty in the latter case are branch cuts that appear due to the specific form of the Coulomb potential.  We present a method for treating branch cuts in quantum-trajectory models, which  is subsequently applied to photoelectron holography. Our method is not numerically intensive, as it does not require finding the location of all branching points and branch cuts prior to its implementation, and is applicable to Coulomb-free and Coulomb-distorted trajectories.  We show that the presence of branch cuts leads to  discontinuities and caustics in the holographic fringes in above-threshold ionization (ATI) photoelectron angular distributions (PAD). These artefacts are removed applying our method, provided they appear far enough from the polarization axis. A comparison with the full solution of the time-dependent Schr\"odinger equation is also performed, and a discussion of the applicability range of the present approach is provided.
	          
\end{abstract}
\maketitle
\section{Introduction}

Above threshold ionization (ATI) 
has allowed unprecedented control in light matter interactions and enabled probing of targets via imaging processes.
One such imaging technique associated with ATI is photoelectron holography \cite{HuismansScience2011,Meckel2014,Haertelt2016}, where, similarly to light holography, an electron freed from a target atom or molecule takes (at least) two paths to the detector: A direct path and one via further interaction with its parent ion.
These paths undergo quantum interference, 
which will yield information about the interaction and binding potential.
Quantum interference is not only vital for photoelectron holography, but plays an important role in molecular imaging \cite{Lein2007} and in the study of other phenomena.
Examples are  high-order harmonic generation (HHG) \cite{Augstein2012}, and, possibly, laser-induced nonsequential double ionization (NSDI) \cite{Faria2008,Hao2014,Maxwell2015,Maxwell2016}.
Besides this, different interpretations of the ionization process, in terms of paths, appeared pivotal for verification of such theoretical concepts as the tunnel exit, the tunneling time and the attoclock setup\cite{Keller2012,Landsman2014,Smirnova2015,Rost2016}.

Traditionally, the strong field approximation (SFA)  \cite{Keldysh1965,Faisal1973,Reiss1980,Popov2004,Popruzhenko2014a} has been widely used for this type of problem.
The power of the SFA is due to the simple analytic solution given for the continuum, which is approximated by field-dressed plane waves.
This, coupled with the ability to build transition amplitudes from many alternative paths to the detector, makes the SFA an ideal method  for the analysis of quantum interference in strong fields.
The specific formulation using the steepest descent method leads to the concept of ``quantum orbits", which are associated with classical orbits and yet may interfere or undergo tunnel ionization \cite{Salieres2001}.
This requires the computation of integrals in complex time, which are straightforward for the SFA.
Also, various representations for quantum-mechanical amplitudes of strong-field processes in terms of other types of paths were proven to be very efficient both in computations and illuminating of the underlying physics; see \cite{Zagoya2012,Zagoya2014,Wu2013,Wu2013b} for a review and references.

Since the past decade, however, it has become clear that the residual Coulomb potential plays an integral role in holographic patterns  \cite{Maharjan2006,Yan2010,Yan2012,Lai2015a,Lai2017,Maxwell2017,Maxwell2017a} observed in photoelectron angular distributions (PADs).
Examples are the fan-shaped structure that forms near the ionization threshold \cite{arbo2008}, the spider-like structure near the field-polarization axis that extends up to high photoelectron energy, and holographic sidelobes \cite{HuismansScience2011,Hickstein2012}.
There have also been other, non-interference based, striking discrepancies between the SFA and ab initio solutions of the TDSE or experiment.
The most significant of these being the low energy (LES) \cite{Blaga2009,Quan2009,Faisal2009}, very low energy (VLES) \cite{Wu2012} and zero energy (ZES) \cite{Wolter2015,Dura2013} structures, respectively.
For that very reason, orbit-based models that include the Coulomb potential,
such as the Eikonal Volkov approximation (EVA) \cite{Smirnova2008}, the time-dependent analytical R-matrix (ARM) method \cite{Torlina2012}, the Coulomb-corrected strong-field approximation (CCSFA) \cite{Popruzhenko2008a} and the Coulomb quantum-orbit strong-field approximation (CQSFA) \cite{Lai2015a} from one side and models of ionization incorporating interference via the semiclassical treatment of the post-tunneling photoelectron dynamics in the two fields \cite{Lein2016} from the other, as well as methods based on a purely classical consideration of the post-ionization dynamics \cite{Rost2012,Rost2016a} have become increasingly popular.
Depending on the model, the influence of the Coulomb potential may be introduced either in the semiclassical action or in the electron trajectories or in both simultaneously.

Nonetheless, in introducing the Coulomb interaction, a number of issues must be faced, namely:
i) The initial conditions of the electron dictate that it must start from the origin, where the Coulomb potential is singular.
This will cause a divergence.
Replacing the singular Coulomb potential by a soft core one does not actually solve this problem, because this does not change the very fact that semiclassical approaches fail in the vicinity of the core.
Thus the necessity to part the space into two subspaces and to match either trajectories or wave functions at the boundary is inevitable for ionization theories with the Coulomb interaction  included \cite{Torlina2013,Zagoya2012}.
ii) The transition amplitude must not be dependent on the integration contour chosen from the complex time of ionization to the real time of detection.
iii) In order to satisfy ii) a calculation based on the trajectory language must necessarily use complex trajectories throughout.
Hence, the Coulomb potential must be extended to the complex-valued position space $\rb$.
As $\sqrt{\rb^2}$ is a multivalued function, this extension raises a question of determining a physically meaningful branch of the square root. 
The problem of choice of ``physical'' leafs for values which are initially determined in the theory as real and then extended into the complex time plane is generally well known in physics, particularly in the theory of scattering \cite{TaylorST}.
Finally, the only real quantity that remains in the theory is the electron momentum, which is measured at the detector.
These issues have been investigated for Coulomb-free trajectory models, where the dynamics are still only determined by the laser field but a phase related to the Coulomb potential is included in the action. Solutions  i), ii) and iii) have been explored in \cite{Perelomov1967,Popruzhenko2008a,Popruzhenko2014b}, \cite{Popruzhenko2014b, Keil2016} and \cite{Popruzhenko2014b, Pisanty2016}, respectively.
Complex Coulomb-free trajectories (i.e. those which solve Newton's equation in the presence of the laser field only, see Section II for details) have also been used in \cite{Torlina2013} in conjunction with the EVA for circularly polarized fields. Therein, it has been shown that the imaginary parts of the trajectories lead to an effective deceleration of the wave packet. This deceleration was found to improve the agreement with ab-initio methods. 

 The methods \cite{Smirnova2008,Popruzhenko2014b} which use Coulomb-free trajectories and correct only the complex-valued Coulomb phase are rigorously justified in the case when the Coulomb interaction only weakly distorts trajectories without changing the topology properties of these trajectories including their mapping onto the final momentum space.
The latter means that, in particular, the number of trajectories bringing the electron into a given final state, the number of close revisits for a given trajectory and other qualitative characteristics which can be expressed in terms of natural numbers do not change under the influence of the Coulomb interaction.
It was however shown that, in many practically interesting cases, the Coulomb interaction considerably changes their topology leading to the emergence of new classes of trajectories which do not exist in the Coulomb-free case \cite{HuismansScience2011}.
These new classes of trajectories were shown to be responsible for the generation of the side lobes, the LES, VLES and other pronounced structures observed in ATI spectra experimentally and in numerical TDSE solutions.
In particular, the Coulomb potential is essential for the holographic patterns to form.
In fact, our previous work \cite{Maxwell2017, Maxwell2017a, Maxwell2018} has shown that Coulomb-distorted trajectories (i.e. those found by solving Newton's equation in the two fields without considering any of them as a perturbation) are required to reproduce the well known fan-like or spider-like structures.
Without Coulomb distortion, the softly forward scattered trajectories involved in producing the spider-like structure do not even exist.
It is however a non-trivial problem to employ complex trajectories if the Coulomb potential is incorporated. 
For that reason, in many Coulomb-corrected approaches the continuum propagation is performed using real variables, which can be justified within the semiclassical picture of the photoionization dynamics \cite{Zagoya2014,Rost2016,Lein2016,Milosevic2017}.
Despite the good agreement between the outcome of such approaches and ab-initio methods, all of them have to consider the ionization step separately, by postulating some initial conditions for the photoelectron emerging at the tunnel exit.
This reduces the applicability of these models to the tunneling limit of ionization and limits the account of non adiabatic effects.
To the best of our knowledge, none of the issues i)-iii) have been dealt with for Coulomb-distorted complex trajectory models.

A method for dealing with  the Coulomb potential energy calculated along a classical trajectory in the complex time plane has been proposed for Coulomb-free trajectories in  \cite{Popruzhenko2014b,Pisanty2016} and later explored in \cite{Keil2016} for a qualitative calculation of photoelectron spectra.
This method requires all branching points and branch cuts to be found and a contour that avoids these branch cuts and remains on a fixed Riemann leaf to be constructed accordingly.
Thus, a different contour must be used for each value of momentum in a PAD.
Additionally, a two-dimensional search through complex time must be performed  and all branch cuts must be mapped in advance. This is a difficult procedure already for Coulomb-free trajectories and may not even be feasible for Coulomb-distorted orbits.

In the present work, we propose a  more efficient method for dealing with branch cuts.
Instead of finding integration contours that avoid branch cuts, we keep the contour fixed by directing it first to the real time axis and then along it.
If a branch cut is encountered along the contour, we take the point for which the branch cut meets the real time axis and the branching point, continue the contour along the cut, circumvent the branch point and return to the real part along the other side of the same cut.
This implies that branch cuts need only be found if they cross that part of the real time axis where the second arm of the integration contour is located, which requires a 1-D rather than 2-D search through complex time.
We apply this method to Coulomb-free and Coulomb-distorted orbits.
Finally, we apply the developed algorithm for a calculation of PADs and show how it can help improving the agreement between the approximate analytic theory and ab initio results.
Apart from bringing a further quantitative progress in the analytic theory of strong-field ionization, our method paves a way to development of an essentially self-consistent method of complex-time quantum orbits which considers the photoelectron interaction with a laser field and with the parent ion on an equal footing.

This article is organized as follows.
In the next section we introduce the basic equations of the theory and specify the two models which deal with Coulomb-free (Subsection A)and Coulomb-distorted (Subsection B) trajectories correspondingly.
Subsection C explains the choice of the integration contour and gives general expressions for the Coulomb-corrected action.
In Section III the contribution of branch cuts into the Coulomb action is calculated and analyzed again for the Coulomb-free (Subsection A) and Coulomb-distorted (Subsection B) cases, and topology of the branching points for different types of photoelectron trajectories is discussed.
Section IV is devoted to a comparative discussion of PADs calculated along different methods.
The concluding section summarizes results and open questions.
We use atomic units throughout.

\section{Background}
The full transition amplitude from a bound state $\left\vert \Psi
_0(t')\right\rangle =\exp[iI_pt']\left\vert \Psi _{0}\right\rangle $ with ionization potential $I_p$  to a final momentum state $ |\tilde{\pb}_f (t)\rangle=|\pb_f+\mathbf{A} (t)\rangle$ reads as \cite{Lai2015a}
\begin{equation}
M(\pb_f)=-i \lim_{t\rightarrow \infty}\hspace*{-0.1cm}
\int_{-\infty }^{t }\hspace*{-0.4cm} \dd t'\hspace*{-0.1cm}
\int \hspace*{-0.1cm}  \left\langle \tilde{\pb}_f (t)
|U(t,t') | H_I(t')| \Psi
_0(t')\right\rangle \,. \label{eq:Mpp}
\end{equation}
In Eq.~(\ref{eq:Mpp}), $U(t,t')$ gives the full time evolution operator
\begin{equation}
U(t,t')=\mathcal{T}\exp \bigg [i \int^t_{t'}H(\tau)d\tau \bigg],
\end{equation}
where $\mathcal{T}$ denotes time-ordering, which relates to the full Hamiltonian
\begin{equation}
H(t)=\frac{\hat{\mathbf{p}}^{2}}{2}+V(\hat{\mathbf{r}})+H_I(t)
\end{equation}
evolving from an initial time $t'$ to a final time $t$. We choose $V(\hat{\mathbf{r}})$ to be the Coulomb potential
\begin{equation}
V(\hat{\mathbf{r}})=-\frac{Z}{\sqrt{\hat{\mathbf{r}}\cdot
		\hat{\mathbf{r}}}},\label{eq:potential}
\end{equation}
with $Z$ being the atomic charge and the interaction $H_I(t)=-\hat{\mathbf{r}}\cdot \mathbf{E}(t)$ with the external field is taken in the length gauge.
Throughout, the hats denote operators. 

In order not to lose generality, we present the theory below without making any assumption on $\mathbf{A}(t)$, except the standard agreement that $\mathbf{A}(-\infty)=\mathbf{A}(+\infty)=0$.
Numerical calculations will be made for the common case of a monochromatic linearly polarized field of frequency $\omega$, with the vector potential
\begin{equation}
\mathbf{A}(t)=\frac{{\bf E}_0}{\omega}\cos\omega t.
\end{equation}
The electric field amplitude $E_0$ is connected to the ponderomotive energy $U_p$ as $U_p=E_0^2/(4\omega^2)$ and the electric field is given by $\mathbf{E}(t)=-d\mathbf{A}(t)/dt$.

\subsection{Coulomb-free Trajectory Model}
\label{sec:Coulfree}
If, in Eq.~(\ref{eq:Mpp}), the binding potential is neglected in $U(t,t')$, the Strong-Field Approximation (SFA) is recovered.
In this case, the drift momentum is conserved and the transition amplitude reduces to
\begin{align}
	M_{\mathrm{d}}=-i\lim\limits_{t\rightarrow\infty}\int_{-\infty}^{t}\hspace*{-0.4cm}\dd t'
	\braket{\tilde{\mathbf{p}}(t)|H_I(t')|\Psi_{0}(t')}e^{i S_\mathrm{d}(\pb,t')},
	\label{Md}
\end{align}
where
\begin{equation}
S_\mathrm{d}(\pb,t')=I_p t'-\int_{t'}^{t}\dd\tau\frac{1}{2}\left(\pb+\Ab(\tau)\right)^2
\label{Sd}
\end{equation}

is the Coulomb-free action associated with the ATI direct electrons.
The time integral in (\ref{Md}) is calculated by the steepest descend method with the well-known expression for the saddle-point equation
$\partial S_{\mathrm{d}}/\partial t'=0$
\begin{equation}
\frac{1}{2}\left[\pb+\Ab(t_s)\right]^2+I_p=0, \label{eq:tunnSFA}
\end{equation}
which can be interpreted  as the conservation of energy at the instant of ionization and has only complex solutions.
The action (\ref{Sd}) calculated at the stationary point can be associated with a trajectory
\begin{equation}
\mathbf{r}_0(\tau)=-\frac{\partial S_d}{\partial{\bf p}}=\int_{t'}^{\tau}\dd \tau'\left(\pb+\Ab(\tau')\right)
\label{eq:r0}
\end{equation}
which starts at the origin at $t=t_s$ and propagates under the action of the laser field only.
Below we refer to such trajectories as {\em Coulomb-free}.
They appear in general as complex-valued, and become real only for the most probable photoelectron momentum minimizing the imaginary part of the action (\ref{Sd}) \cite{Popruzhenko2009}.
Along these trajectories, the Coulomb correction to the action can be calculated in the form
\begin{equation}
S_{\mathrm{C}}(\mathbf{p},t',t)=-\int_{t'}^{t}\dd\tau V[\mathbf{r}_0(\tau)]=
\int_{t'}^{t}\dd\tau\frac{Z}{\sqrt
	{\mathbf{r}_0(\tau)\cdot\mathbf{r}_0(\tau)}	} .
\label{eq:Coulombphase1}
\end{equation}

Within the approximation of Coulomb-free trajectories, the Coulomb effect on the action appear linear with respect to the charge $Z$.
The relative value of the action (\ref{eq:Coulombphase1}) compared to that of the SFA (\ref{eq:tunnSFA}) is given by dimensionless parameters whose value depends on the regime of interaction (see \cite{Popruzhenko2009,Popruzhenko2014b,Popruzhenko2018a}, where such parameters have been introduced in the tunneling and multiphoton regimes of ionization).
Taking the Coulomb distortion of trajectories into account as we explain it in the following, makes the full action a non linear function of $Z$.
Below we consider only hydrogen and therefore set $Z=1$.

The transition amplitude within the saddle-point approximation then reads as
\begin{equation}
M_\mathrm{d}(\pb)=\sum_{s} \mathcal{C}_s(t_s)e^{i S_\mathrm{d}(\pb,t_s)}e^{iS_\mathrm{C}(\rb_0,t_s)},
\label{eq:MpsadCoulfree}
\end{equation}
with
\begin{equation}
\mathcal{C}_s(t_s)=\sqrt{\frac{2\pi i}{{\partial^2S_\mathrm{d}(\pb,t_s)}/{\partial t_s^{2}}}}\braket{\pb+\Ab(t_s)|H_I(t_s)|\Psi_{0}}.
\label{eq:Prefactor}
\end{equation}
Within this approximation there is no need to calculate linear Coulomb corrections to the Coulomb-free trajectories, because their contribution into the action vanishes as shown e.g. in Appendix B of \cite{Popruzhenko2014a}.
Calculation of the prefactor (\ref{eq:Prefactor}) is straightforward for short-range potentials while in the case we are interested in here it requires additional caution.
For potentials with the Coulomb tail, for which the radial part of the bound state wave function behaves at large distances as $R(r)\sim r^{\nu-1}$, where $\nu=1/\sqrt{2I_p}$ is the effective principal quantum number, this calculation is described in details in \cite{Kuchiev1997,Popruzhenko2014a}.

\subsection{Coulomb Quantum-Orbit Strong Field Approximation}

\label{sec:cqsfa}
Inserting the closure relation $\int \tilde{\mathbf{p}}_0\ket{\tilde{\mathbf{p}}_0}\bra{\tilde{\mathbf{p}}_0}=1$ into Eq.~(\ref{eq:Mpp}) gives a transition amplitude in which the electron reaches the continuum via the momentum state $ |\tilde{\pb}_0 (t')\rangle=|\pb_0+\mathbf{A} (t')\rangle$ 
\begin{eqnarray}
M(\pb_f)&=&-i \lim_{t\rightarrow \infty}\hspace*{-0.1cm}
\int_{-\infty }^{t }\hspace*{-0.4cm} \dd t'\hspace*{-0.1cm}
\int \hspace*{-0.1cm} \dd \tilde{\pb}_0 \left\langle \tilde{\pb}_f (t)
|U(t,t') |\tilde{\pb}_0 (t')\right \rangle \hspace*{-0.1cm} \notag  \\
&& \times \left \langle
\mathbf{\tilde{p}}_0 (t')| H_I(t')| \Psi
_0(t')\right\rangle \,. \label{eq:Mpp2}
\end{eqnarray}
The matrix element between the initial and final momentum states enables us to compute Eq.~(\ref{eq:Mpp}) using path-integral methods via time-slicing techniques \cite{Lai2015a}. This gives
\begin{eqnarray}\label{eq:pathMpp}
M(\mathbf{p}_f)&=&-i\lim_{t\rightarrow \infty
}\int_{-\infty}^{t}dt' \int d\mathbf{\tilde{p}}_0
\int_{\mathbf{\tilde{p}}_0}^{\mathbf{\tilde{p}}_f(t)} \mathcal {D}'
\mathbf{\tilde{p}}  \int
\frac{\mathcal {D}\mathbf{r}}{(2\pi)^3}  \nonumber \\
&& \times  e^{i S(\mathbf{\tilde{p}},\mathbf{r},t,t')}
\langle
\mathbf{\tilde{p}}_0 | H_I(t')| \psi _0  \rangle \, ,
\end{eqnarray}
where $\mathcal{D}'\widetilde{\mathbf{p}}$ and $\mathcal{D}\mathbf{r}$ denote integration measures, the prime indicates a restriction and the tilde indicate field-dressed momenta, i.e., $\widetilde{\mathbf{p}}(\tau)=\mathbf{p}+\mathbf{A}(\tau)$. The action in Eq.~(\ref{eq:pathMpp}) reads
\begin{equation}\label{eq:stilde}
S(\mathbf{\tilde{p}},\mathbf{r},t,t')=I_pt'-\int^{t}_{t'}[
\dot{\mathbf{p}}(\tau)\cdot \mathbf{r}(\tau)
+H(\mathbf{r}(\tau),\mathbf{p}(\tau),\tau)]d\tau,
\end{equation}
with
\begin{equation}
H(\mathbf{r}(\tau),\mathbf{p}(\tau),\tau)=\frac{1}{2}\left[\mathbf{p}(\tau)+\mathbf{A}(\tau)\right]^2
+V(\mathbf{r}(\tau)).
\label{eq:Hamiltonianpath}
\end{equation}
Eq.~(\ref{eq:pathMpp}) can also be calculated using saddle-point methods. Minimizing the action (\ref{eq:stilde}) with regard to the intermediate momentum $\mathbf{p}$, the intermediate coordinate $\mathbf{r}$ and the ionization time $t'$ gives
\begin{align}
	\nabla_rS(\mathbf{\tilde{p}},\mathbf{r},t,t')&=\mathbf{0} \implies& \mathbf{\dot{p}}&=-\nabla_rV(\mathbf{r}(\tau)), \label{eq:q-spe} \\\nabla_pS(\mathbf{\tilde{p}},\mathbf{r},t,t')&=\mathbf{0} \implies&
	\mathbf{\dot{r}}&= \mathbf{p}+\mathbf{A}(\tau), \label{eq:p-spe}
\end{align}
and
\begin{equation}
\frac{\left[\mathbf{p}(t')+\mathbf{A}(t')\right]^2}{2}+V(\mathbf{r}(t'))=-I_p,
\label{eq:tunncc}
\end{equation}
respectively.
The latter equation obviously fails for $\mathbf{r}(t')=\mathbf{0}$ for the good reason that the semiclassical approximation employed here does not apply at short distances from the nucleus. For that reason, a series of approximations will be introduced as discussed in Sec.~\ref{subsec:contourchoice}. 
Within the saddle-point approximation, the CQSFA transition amplitude becomes
\begin{equation}
\label{eq:MpPathSaddle}
M(\mathbf{p}_f)\propto-i \lim_{t\rightarrow \infty } \sum_{s}\bigg\{\det \bigg[  \frac{\partial\mathbf{p}_s(t)}{\partial \mathbf{r}_s(t_s)} \bigg] \bigg\}^{-1/2} \hspace*{-0.6cm}
\mathcal{C}(t_s) e^{i
	S(\mathbf{\tilde{p}}_s,\textbf{r}_s,t,t_s)} ,
\end{equation}where $t_s$, $\mathbf{p}_s$ and $\mathbf{r}_s$ are determined by Eqs.~(\ref{eq:q-spe})-(\ref{eq:tunncc}) and $\mathcal{C}(t_s)$ is given by Eq.~(\ref{eq:Prefactor}).  In practice, we use the stability factor $\partial
\mathbf{p}_s(t)/\partial \mathbf{p}_s(t_s)$, which is obtained with a Legendre transformation.
This transformation will lead to the same action and thus not alter the overall dynamics if the electron starts from the origin.
Note that, owing to the Coulomb potential, Eqs.~(\ref{eq:p-spe}),(\ref{eq:tunncc}) are singular at $\mathbf{r}(t')=\mathbf{0}$, so that a series of approximations introduced in the next Subsection are required to make them meaningful.

\subsection{Choice of Contours and Regularization}
\label{subsec:contourchoice}

Actions (\ref{Sd}), (\ref{eq:Coulombphase1}) and (\ref{eq:stilde}) have to be calculated taking the stationary point $t_s$ for the lower integration limit, $t'=t_s$ while the upper integration limit is some real value $t_d$ when the laser field is off and the electron momentum ${\bf p}$ is measured at the detector.
Throughout, unless otherwise stated we consider a two-pronged contour, whose first and second arms are parallel to the imaginary and real time axis, respectively.
The first arm goes from the imaginary time $t'=t'_r+it'_i$ to its real part $t'_r$, while the scond part goes from $t'_r$ to the real final time $t_d$.
This is the most widely used contour in the literature (see, e.g., \cite{Popruzhenko2008a,Yan2012,Torlina2012,Torlina2013}).
In the presence of branch cuts generated by the Coulomb potential energy this contour must be deformed as soon as any of them cross the real axis between $t'_r$ and $t_d$ \footnote{In special cases, this deformation must be applied when the branch cut crosses the tunnel trajectory. This is however a rare occurrence in comparison to it crossing the real axis. }.
The value of the electron coordinate $\mathbf{r}_0(t=t'_r)$ taken at the time $t'_r$ for which the electron reaches the continuum is commonly known as the tunnel exit. 
In the standard version of the CQSFA, we consider only the real part of the tunnel exit. 
This renders all variables in the second part of the contour real, which considerably simplifies the computations.

For the Coulomb-free trajectory model, this will be taken into consideration when computing the trajectory $\mathbf{r}_0$ and the Coulomb phase $S_{\mathrm{C}}(\mathbf{p},t')$ [Eqs.~(\ref{eq:r0}) and (\ref{eq:Coulombphase1}), respectively].
For the CQSFA, the action along the first and the second arm of the contour will be defined as
\begin{eqnarray}
S^{\mathrm{tun}}(\mathbf{\tilde{p}},t'_r,t')&=&iI_pt'_i-\frac{1}{2}\int_{t'}^{t'_r}\left[ \mathbf{p}_0+\mathbf{A} (\tau)\right]^2d\tau \notag \\ &&- \int_{t'}^{t'_r}V(\mathbf{r}_0(\tau))d\tau, \label{eq:stunn}
\end{eqnarray}
and
\begin{eqnarray}
S^{\mathrm{prop}}(\mathbf{\tilde{p}},t,t'_r)&=&I_pt'_r-\frac{1}{2}\int_{t'_r}^{t}\left[ \mathbf{p}(\tau)+\mathbf{A} (\tau)\right]^2d\tau \notag \\&&- \int_{t'_r}^{t}[\mathbf{\dot{p}}\cdot \mathbf{r} +V(\mathbf{r}(\tau))]d\tau, \label{eq:sprop}
\end{eqnarray}
respectively.
In the second arm of the contour, we use
\begin{equation}
\mathbf{r} \cdot \dot{\mathbf{p}}=-\mathbf{r} \cdot \nabla_r V(r)=V(r)
\end{equation}
in Eq.~(\ref{eq:sprop}), which will lead to a factor two in the CQSFA Coulomb phase \cite{Lein2016,Maxwell2017}. Thus, the last
integral in (\ref{eq:sprop}) is equal to
\begin{equation}
\mathcal{I}_C=2\int_{t'_r}^{t}\frac{1}{\sqrt{\mathbf{r}\cdot\mathbf{r}}}d\tau.
\end{equation}

The overall action will then read
\begin{equation}
S(\mathbf{\tilde{p}},t,t')=S^{\mathrm{tun}}(\mathbf{\tilde{p}},t'_r,t')+S^{\mathrm{prop}}(\mathbf{\tilde{p}},t,t_r').
\end{equation}
In the first arm of the contour, we have neglected the influence of the Coulomb  force on the trajectory and have approximated the momentum by $\mathbf{p}_0$. For that reason, Eq.~(\ref{eq:stunn}) is given in terms of $\mathbf{p}_0$ and $\mathbf{r}_0$. In this case, Eq.~(\ref{eq:tunncc}) is approximated by
\begin{equation}
\frac{1}{2}\left[ \mathbf{p}_0+\mathbf{A}(t')\right]^2+I_p=0.
\label{eq:tp-spe}
\end{equation}
Although this equation is formally the same as (\ref{eq:tunnSFA}), the solutions will be different as it will be matched to the solutions of saddle-point equations (\ref{eq:p-spe}) and (\ref{eq:q-spe}) describing the electron motion in the continuum.
Physically, Eq.~(\ref{eq:stunn}) relates to the sub-barrier dynamics, while Eq.~(\ref{eq:sprop}) is associated with the continuum propagation.

When the integrand remains an analytic function everywhere in the complex time plane except the infinity, as is the case for the Coulomb-free action (\ref{Sd}), the integration contour connecting $t_s$ and $t_d$ can be chosen arbitrarily.
This makes, in particular, the value of tunnel exit ill-defined  for theories where the Coulomb field is fully discarded \cite{Popruzhenko2014a,Popruzhenko2014b}.
In contrast, the integrands in (\ref{eq:Coulombphase1}) and (\ref{eq:stilde}) may have singularities generated by the divergency of the Coulomb potential energy at the origin and at those points where ${\bf r}^2=0$.
As  ${\bf r}(t)$ is a complex-valued vector, the condition ${\bf r}^2=0$ does not necessarily mean that ${\bf r}=0.$
This imposes certain constraints on the choice of the integration contour \cite{Popruzhenko2014a,Popruzhenko2014b,Pisanty2016,Keil2016}.
Besides the integrals (\ref{eq:Coulombphase1}), (\ref{eq:stunn}) and (\ref{eq:sprop}) diverge logarithmically at both endpoints of the contour, i.e., at the initial time $t'$ and the final time $t_d \rightarrow \infty$, due to the presence of the Coulomb term.
However, as the asymptotic behavior of the Coulomb integral in the limit $t_d\to\infty$ is the same for all trajectories corresponding to the same final momentum, $S_C({\bf p},t)\sim p^{-1}\ln t$, and this divergent contribution is real, the upper-limit divergency does not influence the shape of photoelectron spectra.
Instead, the divergency at the lower integration limit is generated by the Coulomb singularity and physically connected to the fact that the approximate treatment of the Coulomb interaction along a Coulomb-free trajectory does not apply when the electron approaches the atomic core, so that the Coulomb interaction becomes dominant.

To overcome this issue, we follow the regularization procedure outlined in  \cite{Popruzhenko2014a,Popruzhenko2014b}, in which the Coulomb phase (\ref{eq:Coulombphase1}) is matched to the asymptotic value for the bound-state Coulomb wave function $ \Psi_{at}(t_{\rm m})$ at a time $t_{\rm m}$ such that $1/\sqrt{2I_p}\ll {\bf r}(t_{\rm m})\ll E_0/\omega^2$, so that the electron is already sufficiently away from the atom, but still travelled a small fraction of its quiver amplitude.
The generally complex-value matching point satisfying these conditions can be arbitrarily chosen around $t_s$; as it vanishes from the final result there is no need to specify its precise position on the integration contour.

In both cases of the Coulomb-free trajectories and of the CQSFA the regularized integral along the vertical arm of the contour takes the form \cite{Perelomov1967,Popruzhenko2009,Popruzhenko2014a}:
\begin{equation}
S_C=-i\nu\ln\bigg\lbrack 2I_pt_i^{\prime}\bigg\rbrack-i\int\limits_0^{t_i^{\prime}}\bigg(\frac{Z}{\sqrt{{\bf r}_0^2(t_r^{\prime}+i\tau)}}-\frac{\nu}{t_i^{\prime}-\tau}\bigg)d\tau
\end{equation}
After the regularization is performed, the remaining integral along the real axis preserves the form it has in Eqs.~(\ref{eq:Coulombphase1}) and (\ref{eq:sprop}) correspondingly with the difference that in the integral along a Coulomb-free trajectory the lower integration limit is replaced by $t_r^{\prime}$.
Although the Coulomb integrals along the second arm look identical in both approaches, they are nevertheless different: a Coulomb-free trajectory acquires a constant imaginary part, generating branch points and first-order poles, which should be accounted for during the integration.
In the formulation of the CQSFA given in \cite{Lai2015a,Lai2017,Maxwell2017}, the trajectories were assumed real along the real time contour, so that the integration process is straightforward.
This assumption will be relaxed in the next sections.\section{Treatment of branch cuts}
\label{sec:BCcontours}

The solutions of the saddle-point equations in Secs.~\ref{sec:Coulfree} and \ref{sec:cqsfa} are complex, so that the corresponding trajectories $\rb(t)$ are complex-valued as well.
The only observable, and therefore guaranteed real quantity, is the final  momentum $\mathbf{p}_f(t_d)$, $t_d \rightarrow \infty$ at the detector. Thus, the residual binding potential
\begin{equation}
V[\rb(\tau)]=-\frac{1}{\sqrt{\rb(\tau)\cdot\rb(\tau)}}
\end{equation}
must be extended to the complex plane where it exhibits branch cuts \cite{Popruzhenko2014a,Popruzhenko2014b,Pisanty2016,Keil2016}. If the standard convention is applied, branch cuts occur in the negative real half axis of $\rb(\tau)\cdot\rb(\tau)$, i.e., for
\begin{align}
	\Re\left( \rb(\tau)\cdot\rb(\tau) \right)<0 && \mathrm{and} \qquad\Im\left( \rb(\tau)\cdot\rb(\tau) \right)=0.
	\label{eq:Branchcut}
\end{align}
Hence, the integration contour used must not cross this line.
In this section, we present an alternative solution to that adopted in  \cite{Keil2016,Pisanty2016} to treat branch cuts. Instead of mapping them in advance and using this information to build a specific contour, we first choose a contour and, after computing the trajectory, test to see if any branch cut has been crossed. Should this be the case, then the contour must be deformed in such a way  that it goes around the branch cut. 
Hence, one must find the branching point in complex time, which will be defined  by
\begin{align}
	\rb(\tau)\cdot\rb(\tau) = 0+0 i.
	\label{eq:BranchEnd}
\end{align}
Given a time $t_b$ along the second arm of the contour that satisfies Eqs.~(\ref{eq:Branchcut}), it is easy to use that to find a complex time $t_k$ that satisfies Eq.~(\ref{eq:BranchEnd}).
Throughout, we will use the two-pronged contour specified in Sec.~\ref{subsec:contourchoice}, whose first and second arms are chosen parallel to the imaginary axis and along the real time axis, respectively.
For this specific contour, $t_b$ will mark the intersection of the branch cut with the real axis. Together with the time $t_k$ marking the end of the branch cut, this can be used to  construct a contour from three parts that can be integrated over to `correct' the potential integral around each branch cut.
The Coulomb phase is then computed along this contour, so that
\begin{align*}
	S^{(cut)}_{C}=\int_{c_1}V[\rb(\tau)]\dd \tau+\int_{c_2}V[\rb(\tau)]\dd \tau+\int_{c_3}V[\rb(\tau)]\dd \tau,
\end{align*}
where $c_1$ goes from $t_b$ to $t_k$ following the branch cut, $c_2$ goes back from $t_k$ to $t_b$ and $c_3$ connects them together with a half circle around $t_k$. By taking a function
\begin{equation}
u_{\pm}=\mathbf{r}(\tau) \cdot \mathbf{r}(\tau)=|u(\tau)|\exp(\pm i\pi),
\end{equation}
such that $u_{+}$ is restricted to $c_1$ and $u_{-}$ to $c_2$, one may easily show that
\begin{equation}
\int_{c_1}V[\rb(\tau)]\dd \tau=\int_{c_2}V[\rb(\tau)]\dd \tau
\end{equation}
and that, for $p_{\perp}\neq 0$, the integral over $c_3$ vanishes. For $p_{\perp}=0$ the topology of the problem changes, with poles instead of branch cuts, and there will be divergencies. 
We will discuss this problem in the subsequent sections, see also \cite{Popruzhenko2018b}.

\begin{figure*}
	\includegraphics[width=\textwidth]{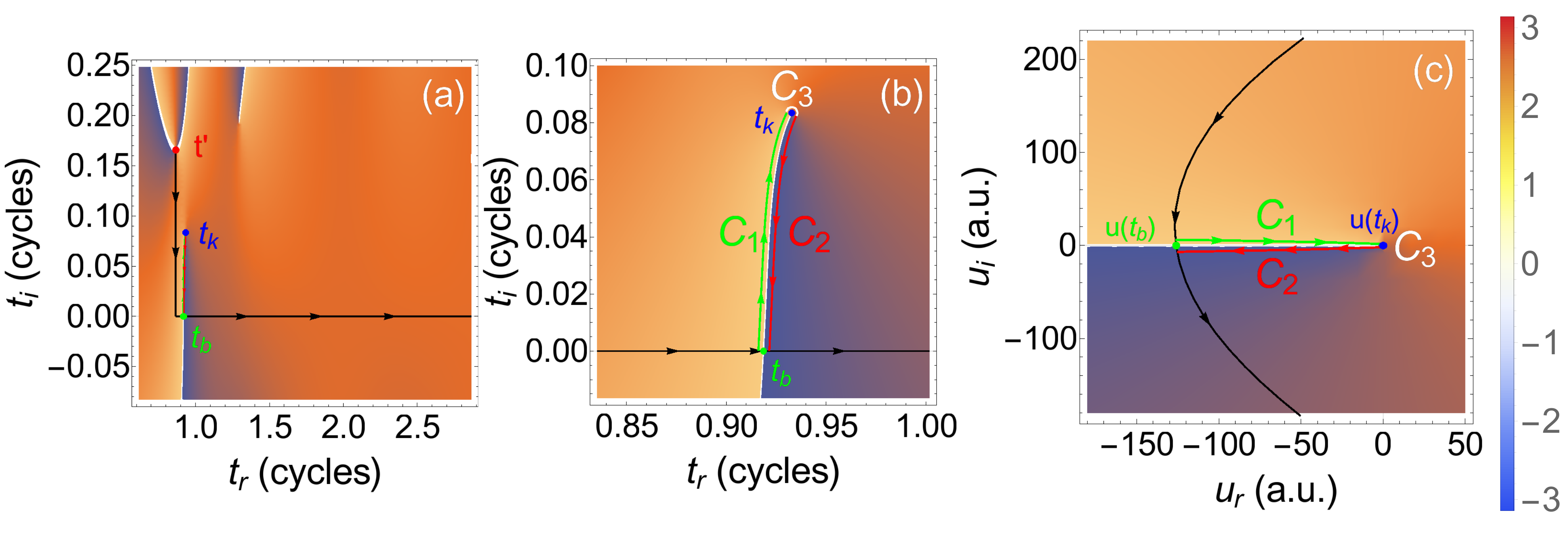}
	\caption{Example of a contour constructed around a branch cut according to the procedure stated in this section, using Coulomb-free trajectories. Panel (a) shows the standard contour in the complex time plane, together with the branch cuts obtained by plotting $\arg(\sqrt{\mathbf{r}_0(\tau)\cdot\mathbf{r}_0(\tau)})$. Panel (b) displays the close up of the branch cut and the contour around it. Panel (c) provides a view of the contour in terms of the real and imaginary parts of $\rb \cdot \rb$. We have chosen momentum components $(p_{f\parallel},p_{f\perp})=(-1.4 \hspace*{0.1cm}\mathrm{a.u.}, 0.7\hspace*{0.1cm}\mathrm{a.u.})$. The field and atomic parameters are $I_p=0.5$ a.u,    $\omega = 0.0570$ a.u. and  $U_p = 0.439$ a.u.    ($\lambda = 800$ nm, $I= 2\times 10^{14} \mathrm{W/cm}^2$).}
	\label{fig:BranchDiagram}
\end{figure*}

For a schematic representation of this contour and how it is deformed to avoid branch cuts, see Fig.~\ref{fig:BranchDiagram}. Branch cuts manifest themselves as discontinuities in the argument of $\sqrt{\mathbf{r}(\tau)\cdot \mathbf{r}(\tau)}$, and thus may be visualized if this argument is plotted in the complex time plane.

A key question is whether any electron orbit given by a solution of the saddle-point equations crosses one or more branch cuts, either for the Coulomb-free case or for the CQSFA, and how this depends on the momentum components parallel and perpendicular to the polarization axis. For a monochromatic field, specific saddle-point solutions are given by
\begin{align}
	t_{a} &= \frac{2\pi}{\omega}-\frac{1}{\omega} \arccos\left(\frac{-p_{0\parallel}+ i\sqrt{2 I_\mathrm{p}+p^2_{0\perp}}}{2\sqrt{U_\mathrm{p}}}\right), \label{eq:t1s}\\
	t_{b}&= \frac{1}{\omega} \arccos\left(\frac{-p_{0\parallel}- i\sqrt{2 I_\mathrm{p}+p^2_{0\perp}}}{2\sqrt{U_\mathrm{p}}}\right) \label{eq:t2s}
\end{align}
where $p_{0\parallel}$ and $p_{0\perp}$ are the components of the initial momentum parallel and perpendicular to the field polarization axis, and $a,b$ refer to types of orbits that will depend on the model. 

\subsection{Coulomb-free trajectories}
\label{subsec:restrajsSFA}

\begin{figure}
	\includegraphics[width=0.5\textwidth]{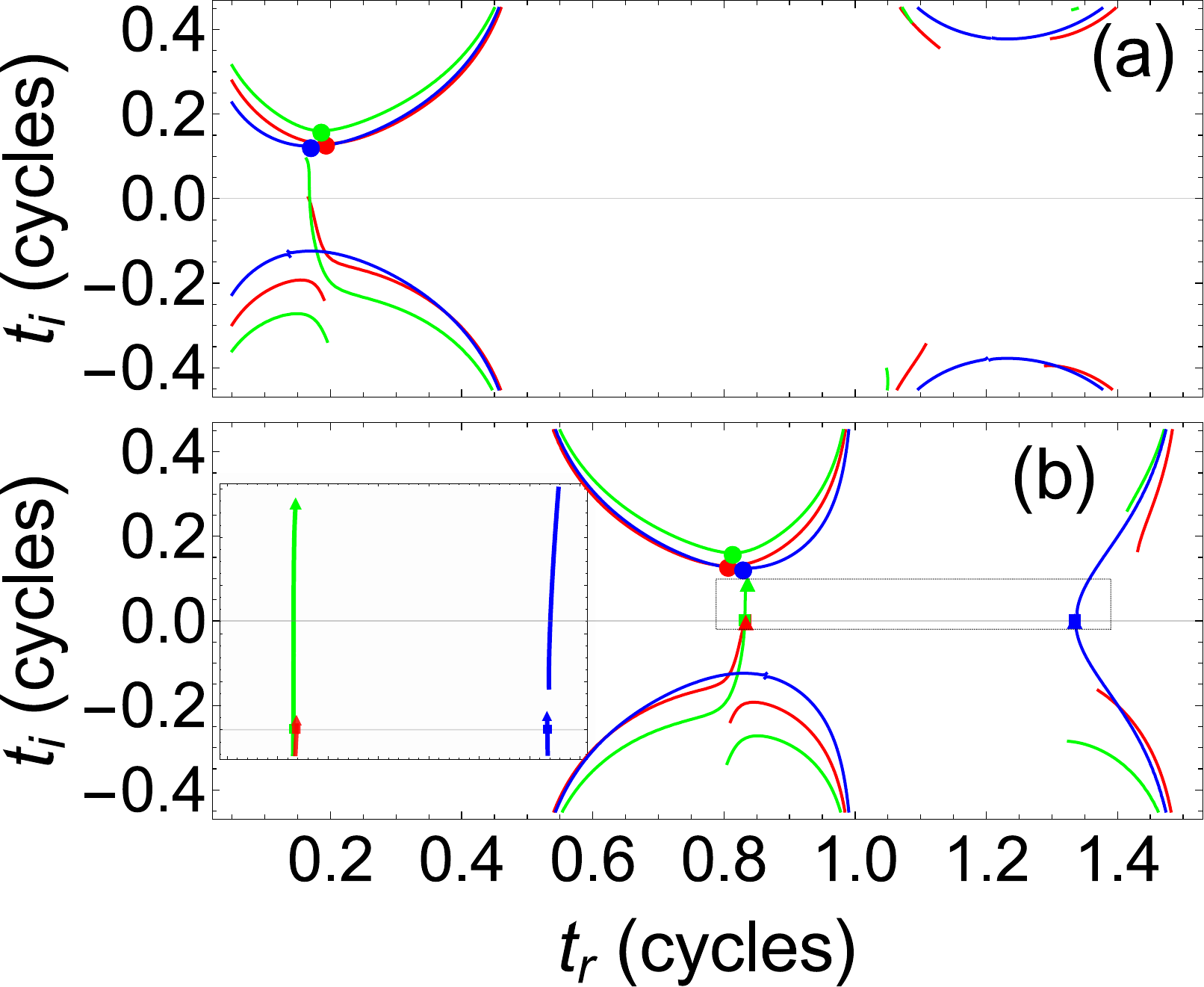}
	\caption{Branch cuts in the complex time plane calculated for Coulomb-free trajectories corresponding to three different final momenta and the same field and atomic parameters as in the previous figure. The dots in the figure indicate the start $t'$ of the time contour for each set of parameters, according to Eqs.~(\ref{eq:t1s}) and (\ref{eq:t2s}) (orbits 1 and 2, respectively), the squares give the time $t_b$ at which a branch cut crosses the real time axis provided $t_b>t'_r$, and the triangles mark the branching points $t_k$. Panel (a) refers to orbit 1, while panel (b) refers to orbit 2. A blow up of the region where the branch cuts meet the real axis for physically relevant parameters is provided on the left hand side of panel (b). The red, green and blue curves in the figure correspond to the momentum components  $(p_{f\parallel},p_{f\perp})=(-0.63, 0.53)$,     $(p_{f\parallel},p_{f\perp})=(-0.80, 1.05)$   and  $(p_{f\parallel},p_{f\perp})=(-0.82, 0.01)$ given in atomic units.  }
	\label{fig:Coulombfree}
\end{figure}
For Coulomb-free trajectories, the complex coordinate $\mathbf{r}_0(\tau)$ [Eq.~(\ref{eq:r0})] may be computed analytically and one may easily map the branch cuts.  For $p_{\parallel}>0$, the ionization times given by Eqs.~(\ref{eq:t1s}) and (\ref{eq:t2s}) are associated to orbits 1 and 2 according to the classification in \cite{Yan2010}. An electron along orbit 1 is freed in the direction of the detector, while for orbit 2 it is initially released in the opposite direction. For $p_{\parallel}<0$, the situation is reversed and the solutions are shifted by half a cycle. One should note that in the Coulomb-free case the drift momentum is conserved so that $\mathbf{p}_{0}=\mathbf{p}_{f}\equiv\mathbf{p}$.

Fig.~\ref{fig:Coulombfree} shows an example of a branch-cut mapping for orbits 1 and 2 (upper and lower panels, respectively). The figure shows two sets of branch cuts, in the lower and upper complex time plane, whose shape and distance from the real time axis depends on the electron momentum components, and on the field and atomic parameters. The momentum components  used in Fig.~\ref{fig:Coulombfree} have been carefully chosen to match clear features in the ATI photoelectron angular distributions, which will be addressed in the next Section. 

For orbit 1 [panel (a)], the branch cuts are in general away from the real time axis, unless $\mathrm{Re}[t]$ is smaller than the classical ionization times. These solutions are not relevant as we have defined the contour such that ${\rm Re}(t)\ge t_r$ along it.
For orbit 2, the situation is quite different, and the branch cuts intersect the real time axis for physically relevant times. This means that they must be taken into consideration. 
The red lines correspond to a region in which our chosen contour meets a branch cut, and for which our method is successfully applied.  

There are clear gaps between the upper and lower sets of branch cuts. This is even more visible for larger values of $t_r$, for which the gaps are wide and the branch cuts are located far from the real time axis.
The green curve is also in a region for which the corrections can be successfully applied. The main difference in this case is the perpendicular momentum $p_{\perp}$, which was chosen to be quite large. This leads to longer branch cuts, and the branching points move away from the real time axis. This can be clearly seen for the branch cut and branching point close to the start time $t'$. The gaps between the lower and upper sets of branch cuts also widen. Finally, the blue curves illustrate a case for which the branch-cut corrections do not work, namely small scattering angles. In this case, the gaps between the two different sets of branch cuts are quite small or even close, so that the contour can no longer be deformed as discussed. 
For decreasing perpendicular momentum, the gap between the two branching points becomes increasingly shorter until, for $p_{\perp}=0 $ they merge into a pole \cite{Popruzhenko2018b} leading to divergencies.

Eqs.~(\ref{eq:Branchcut}) defining a branch cut can also be employed for finding regions in position space along the chosen contour for which branch cuts occur. For a Coulomb free orbit, we find, for a time $\tau_r$ such that $t'_r<\tau_r<t$ lies along the real axis, that
\begin{equation}
|\mathbf{r}_{0r}(\tau_r)|^2<|\mathbf{r}_{0i}|^2 \qquad  \mathrm{and} \qquad \mathbf{r}_{0r}(\tau_r)\cdot \mathbf{r}_{0i}=0,
\label{eq:rbranchcutSFA}
\end{equation}
where $\mathbf{r}_{0r}(\tau_r)$ and $\mathbf{r}_{0i}$ give the real and imaginary parts of the Coulomb-free trajectory (\ref{eq:r0}). 
One should note that the imaginary part of this orbit is constant and given by the phase picked up at the instant of ionization. This leads to two straight segments starting radially from the origin, which have a length of $|\mathbf{r}_{0i}|$. If these segments  are crossed by a specific orbit, then this orbit is crossing a branch cut.
\begin{figure}
	\includegraphics[width=0.5\textwidth]{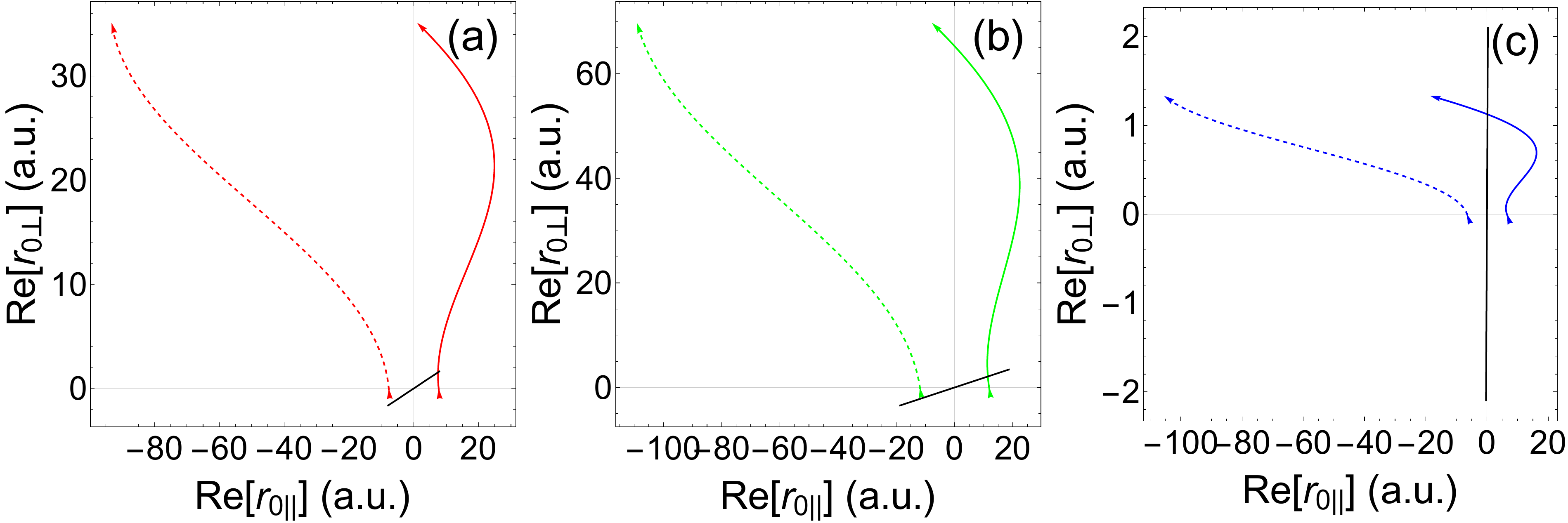}
	\caption{Orbits 1 and 2, whose ionization times are given by the solution of Eqs.~(\ref{eq:t1s}) and (\ref{eq:t2s}) computed for momentum components $(p_{f\parallel}, p_{f\perp})=(-0.63, 0.53)$, $(p_{f\parallel},p_{f\perp})=(-0.80, 1.05)$   and  $(p_{f\parallel},p_{f\perp})=(-0.82, 0.01)$ in atomic units [panels (a), (b) and (c), respectively], together with the condition (\ref{eq:rbranchcutSFA}), which is illustrated as the black segments in the figure.  We consider the tunnel exit $z_0$ such that its real part is larger than zero. The field and atomic parameters are the same as in the previous figure.}
	\label{fig:branchSFAorb}
\end{figure}
Fig.~\ref{fig:branchSFAorb} displays these conditions together with the real part of the coordinate $\mathbf{r}_0$ obtained for the two solutions of Eqs.~(\ref{eq:t1s}) and (\ref{eq:t2s}) that exist in the Coulomb-free case. The figure shows that orbit 1 does not meet any branch cut, while orbit 2 does.  
One should note that, for the Coulomb-free trajectories, the segment in the figure gives condition (\ref{eq:rbranchcutSFA}) for both orbits 1 and 2.

\subsection{Coulomb-distorted trajectories}
\label{subsec:restrajsCQSFA}

For Coulomb-distorted trajectories in the framework of the CQSFA, the task of mapping  is quite involved, as the continuum propagation requires solving the coupled ordinary differential equations (\ref{eq:q-spe}), (\ref{eq:p-spe}) numerically in the complex plane using the initial conditions given by the complex  tunnel exit $\mathbf{r}_0(t'_r)$. The key difficulty is that these equations contain branch cuts themselves. 
In order to avoid this problem, in our previous work \cite{Maxwell2017, Maxwell2017a, Maxwell2018} the imaginary part of the tunnel exit was discarded, which led to real variables in the second arm of the contour. This is however an oversimplification, as the electron trajectories are complex throughout.

To make this problem tractable, we retain the imaginary part of the trajectories, but assume it behaves as in the Coulomb-free case in the continuum propagation. This simplification relies on the assumption that the imaginary parts do not differ too much from their Coulomb-free counterparts as (i) they stem mainly from the sub-barrier dynamics; (ii) within the CQSFA, the sub-barrier dynamics are described in the same way as for the Coulomb-free case, in the sense that the momentum is kept constant and the coordinate is given by $\mathbf{r}_0(t)$. We also assume that, along the real time axis, the imaginary part of the trajectory is constant and equal to the imaginary part of the field-free trajectory at the tunnel exit. This gives\begin{equation}
\mathbf{r}_{b}(\tau_r)=\mathbf{r}(\tau_r)+i \mathbf{r}_{0i},\label{eq:rapproxaxis}
\end{equation}
where $\mathbf{r}_{0i}=\mathrm{Im}[\mathbf{r}_{0}(t_r')]$. This simplification loosely relates to the fact that, in the limit $t \rightarrow \infty$, the imaginary part of $\mathbf{r}_b(t)$ must be constant so that the final momentum $\mathbf{p}_f$ at the detector is real.

Away from the real time axis, we consider
\begin{equation}
\mathbf{r}_{b}(\tau)=\mathbf{r}(\tau_r)+i \mathbf{r}_{0i}+\mathbf{r}_c(\tau),\label{eq:rapprox}
\end{equation}
where
\begin{equation}
\mathbf{r}_{c}(\tau)=\int_{\tau_r}^{\tau}\mathrm{d}\tau'\left(\mathbf{p}(\tau'_r)+\mathbf{A}(\tau')\right).
\end{equation}
to be able to treat branch cuts.

The real parts of the solutions are computed as previously, using the real part of the tunnel exit and the solutions (\ref{eq:t1s}), (\ref{eq:t2s}) as initial conditions, for a given final momentum $\mathbf{p}_f$. In the Coulomb-corrected case, there will be four types of solutions, whose classification is discussed in detail in \cite{Yan2010}. For initial parallel momentum $p_{0\parallel}>0$, orbits type 1 and 4 start in the direction of the detector and are determined by the ionization time (\ref{eq:t2s}), and orbits type 2 and 3 start in the opposite direction at times given by Eq.~(\ref{eq:t1s}). For $p_{0\parallel}<0$ the situation is reversed. Each of these orbits is however topologically different: An electron along orbit 1 will go directly to the detector, while an electron along orbit 4 will first go around the core. For orbit 2, the electron's perpendicular momentum will not change its sign, while $p_{0\perp}p_{f\perp}<0$ for orbit 3.

\begin{figure}
	\includegraphics[width=0.5\textwidth]{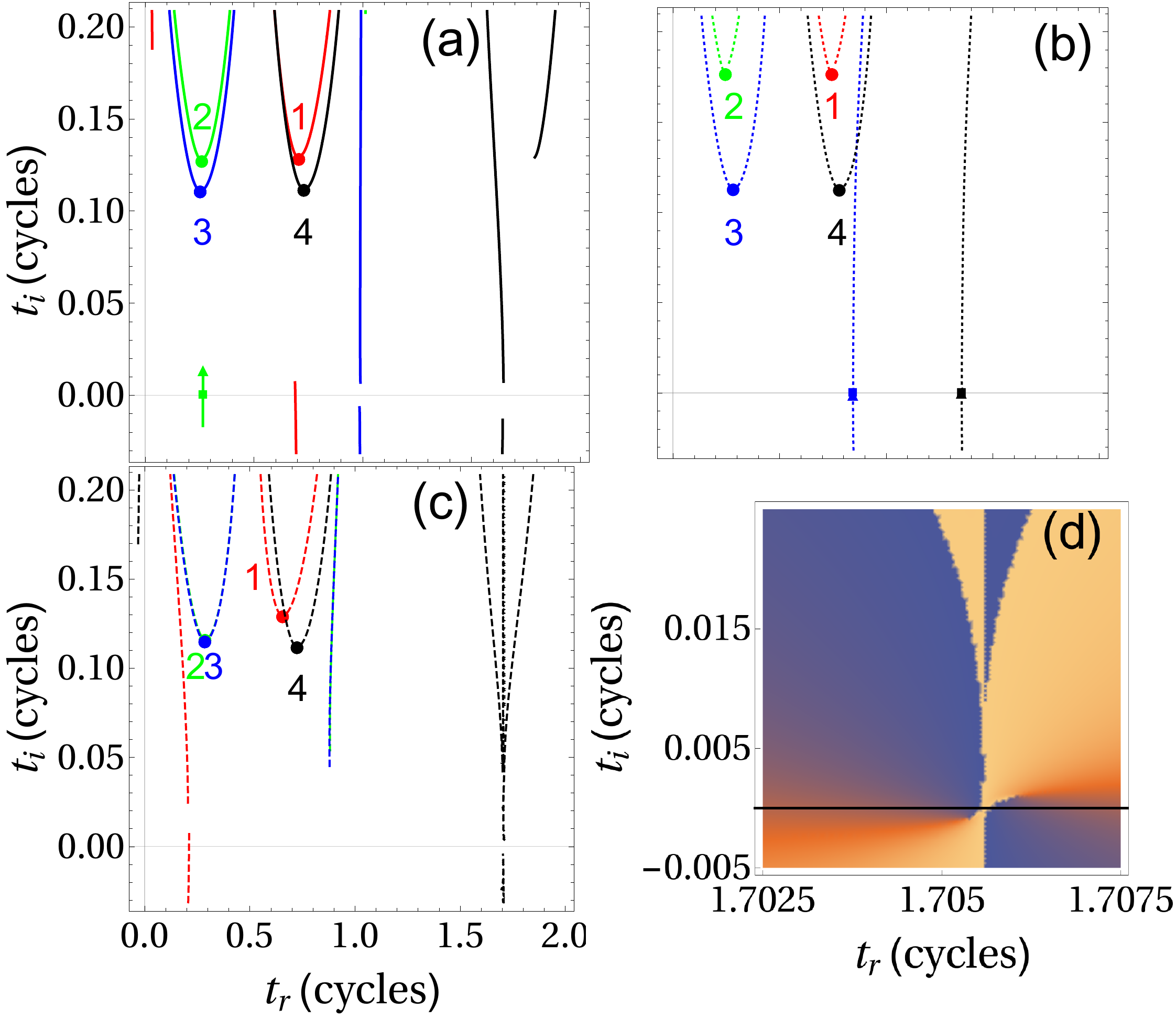}
	\caption{Branch cuts in the complex time plane for different, fixed momenta, Coulomb-distorted trajectories and the same field and atomic parameters as in the previous figure. In panels (a), (b) and (c), we have used the momentum components $(p_{\parallel},p_{\perp})=(-0.475, 0.400) $,   $(p_{\parallel},p_{\perp})=(-0.604,0.980) $  and $(p_{\parallel},p_{\perp})=(-0.619,0.0113)$, respectively.  The momenta are given in atomic units. The branch cuts related to orbits 1, 2, 3 and 4 are displayed as the red, green, blue and black lines in the figure. The corresponding ionization times $t'$, branching times $t_k$ and the intersection times $t_b$ of the branch cut with the real time axis are illustrated as dots, triangles and squares  using the same color convention. 	For clarity, the orbit number is indicated close to the ionization times. In different panels, we have used solid, dotted or dashed lines to match the trajectories employed in Fig.~\ref{fig:branchCQSFAorb}. In panel (d), we plot $\arg(\sqrt{\mathbf{r}_b(\tau)\cdot\mathbf{r}_b(\tau)})$ computed according to Eq.~(\ref{eq:rapprox}) for the parameters in panel (b), in the vicinity of the overlapping branch cuts. The remaining field and atomic parameters are the same as in the previous figures.}
	\label{fig:branchmapCQSFA}
\end{figure}

Fig.~\ref{fig:branchmapCQSFA} illustrates the branch cuts in the complex plane obtained from the discontinuities of $\arg(\sqrt{\mathbf{r}_b(\tau)\cdot\mathbf{r}_b(\tau)})$, where $\mathbf{r}_b(\tau)$ has been computed according to Eq.~(\ref{eq:rapprox}) for orbits 1, 2, 3 and 4. The overall behavior is similar to that in the Coulomb-free case, i.e., two sets of branch cuts separated by gaps, which, depending on the parameters chosen, approach or distance themselves from the real time axis. In Fig.~\ref{fig:branchmapCQSFA}(a), we have chosen the momentum components so that, for the contour used, a branch cut will be crossed for orbit 2 (see green line therein). For the remaining orbits, branch cuts are avoided. This is due to the fact that $\mathrm{Re}[t']$ is smaller than $t_b$ for orbit 1, and that there are clear gaps between the two sets of branch cuts that exist for orbits 3 and 4. The branch cut crossing orbit 2 will lead to fringe discontinuities in holographic patterns dependent on this specific orbit, such as the fan and the spider. 

In Fig.~\ref{fig:branchmapCQSFA}(b), the final momentum components were taken so that, for a given energy, the scattering angle of orbit 3 is maximized. This gives a momentum very close to the $p_{\perp}$ axis. The figure clearly shows that the branch cuts intersect the $t_r$ axis for orbits 3 and 4. This is associated with acts of rescattering. The Coulomb-distorted trajectories tend to behave in a more complex way than their Coulomb-free counterparts. 
For instance, they may even cross the branch cuts twice within a fraction of a field cycle. An example of that is provided for orbit 4 in  Fig.~\ref{fig:branchmapCQSFA}(c). The figure shows that, near the $p_{f\parallel}$ axis, two or more branch cuts may overlap, which renders the present algorithm inapplicable (see black line therein). It is important to stress that there is not a simple gap between non-overlapping branch cuts as in the previously discussed cases, as $\arg(\sqrt{\mathbf{r}_b(\tau)\cdot\mathbf{r}_b(\tau)})$ clearly exhibits many discontinuities in this region. For clarity, we have plotted this argument near the branch-cut overlap in Fig.~\ref{fig:branchmapCQSFA}(d).

\begin{figure}
	\includegraphics[width=0.5\textwidth]{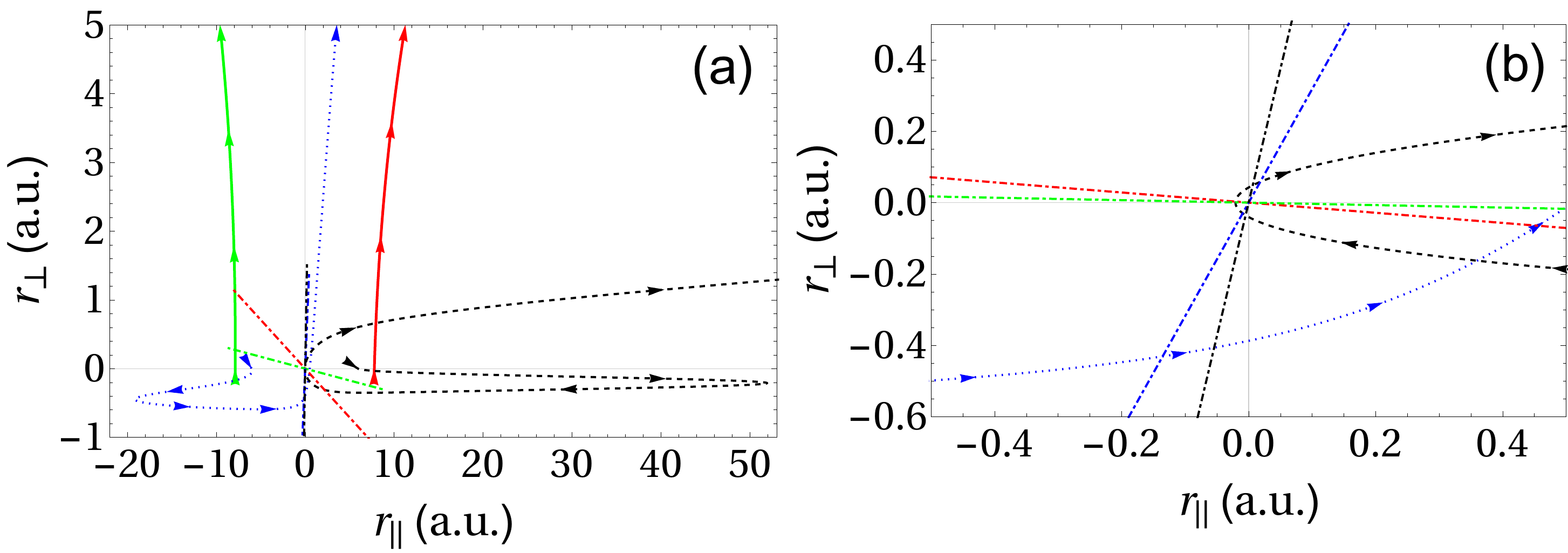}
	\caption{Orbits 1 to 4 computed for the same field and atomic parameters as in Fig.~\ref{fig:branchmapCQSFA}, together with the conditions (\ref{eq:rbranchcutCQSFA}). The solid, dotted and dashed lines correspond to  $(p_{\parallel},p_{\perp})=(-0.475, 0.400) $,   $(p_{\parallel},p_{\perp})=(-0.604,0.980) $  and $(p_{\parallel},p_{\perp})=(-0.619,0.0113)$, respectively.   The momenta are given in atomic units. The color convention employed is the same as in the previous figure, i.e., the red, green, blue and black lines yield orbits 1, 2, 3 and 4, respectively. Panel (a) provides an overall view of the orbits, while panel (b) gives a blow up near the core in order to illustrate rescattering events and multiple passes.}
	\label{fig:branchCQSFAorb}
\end{figure}

Similarly to what has been done in the Coulomb-free case, it is possible to use Eqs.~(\ref{eq:Branchcut}) to define conditions upon $r_b(\tau_r)$, where $t'_r<\tau_r<t$ is real, in order to determine whether the present contour meets a branch cut.  Using Eq.~(\ref{eq:rapprox}) on the real time axis, we obtain
\begin{equation}
|\mathbf{r}(\tau_r)|^2<|\mathbf{r}_{0i}|^2 \qquad  \mathrm{and} \qquad \mathbf{r}(\tau_r)\cdot \mathbf{r}_{0i}=0.
\label{eq:rbranchcutCQSFA}
\end{equation}
Fig.~\ref{fig:branchCQSFAorb} illustrates these conditions, for orbits 1, 2, 3 and 4. The figure shows that, for orbits 3 and 4, the trajectories crossing a branch cuts can be directly related to rescattering events. This is in agreement with the fact that, under some circumstances, these orbits may be identified with SFA rescattered orbits \cite{Maxwell2018}. A blow up of the figure near the core clearly shows the hard collision that occurs for orbit 3 mentioned in the discussion of Fig.~\ref{fig:branchmapCQSFA}(b), and the double pass near the core related to the overlapping branch cuts in Figs. Figs.~\ref{fig:branchmapCQSFA}(c) and (d). For clarity, we have used the same line styles in both figures. 

\section{Photoelectron angular distributions}
\begin{figure}
	\includegraphics[width=0.5\textwidth]{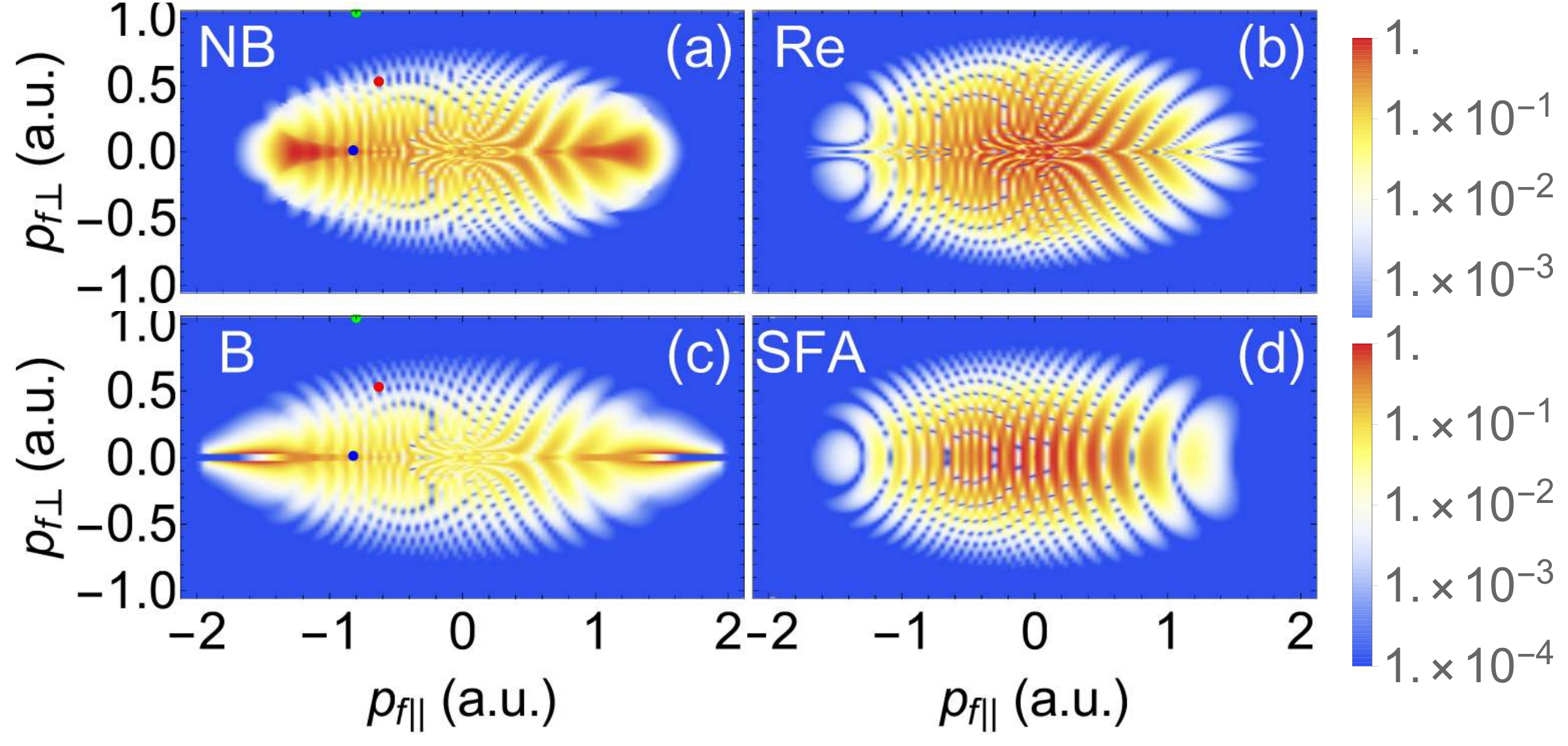}
	\caption{Photoelectron angular distributions computed using Coulomb-free trajectories for Hydrogen ($I_p=0.5$ a.u.) in an field of $\omega = 0.0570$ a.u., and $I= 2 \times 10^{14} \mathrm{W}/\mathrm{cm}^2$  ($\lambda = 800 $nm,  $U_p$ = 0.439Â Â Â). In panels (a) to (c), we have included  the Coulomb phase (\ref{eq:Coulombphase1}) and Coulomb-free trajectories, while in panel (d) the outcome of the plain SFA is displayed for comparison. In Panels (a) and (c), we considered complex trajectories without and with branch-cut corrections, while in panel (b) real trajectories have been used.   The red, green and black dots in panels (a) and (c) correspond to momentum components  $(p_{f\parallel},p_{f\perp})=(-0.63, 0.53)$    $(p_{f\parallel},p_{f\perp})=(-0.80, 1.05)$   and  $(p_{f\parallel},p_{f\perp})=(-0.82, 0.01)$ given in atomic units. All plots have been displayed in a logarithmic scale.  }
	\label{fig:padcoulfree}
\end{figure}

In Fig.~\ref{fig:padcoulfree}, we show the ATI PADs obtained using Coulomb-free trajectories of several types [panels (a) to (c)] and the Coulomb-corrected action, together with the plain SFA [panel (d)]. If complex trajectories are used without branch cut corrections, there is an anomalous trumpet-shaped structure near the $p_{\parallel}$ axis, and a fringe dislocation at larger transverse momenta [Fig.~\ref{fig:padcoulfree}(a)]. 
We have verified that both features are directly related to orbit 2 crossing branch cuts. 
The different Riemann sheets can indeed be related to two elliptical regions in momentum space, centered at nonvanishing perpendicular momenta and vanishing parallel momenta. If the branch cuts are corrected as described in the previous section [Fig.~\ref{fig:padcoulfree}(c)], both the fringe discontinuity and the trumpet shape structure vanish and a PAD extending to much higher parallel momenta is obtained. 
The discontinuity along the $p_{\parallel}$ axis is related to the fact that the pairs of branching points merge into a pole when $p_{\bot}\to 0$, and the contributions of the segments $C_1$ and $C_2$ of the integration contour shown on Fig.\,1 become logarithmically divergent. 
This distribution is very distinct from that obtained using real Coulomb-free trajectories, which is displayed in Fig.~\ref{fig:padcoulfree}(b). 
Using real trajectories seem to overestimate the width of the PAD with regard to the perpendicular momentum $p_{\perp}$. 
This agrees with the statement in \cite{Torlina2013} that the imaginary components of the continuum trajectories decelerates the electron wavepacket. 
A visible extension of the distributions towards higher values of $p_{\parallel}$ agrees qualitatively with the results of \cite{Keil2016}, where an order in magnitude enhancement in the probability of ionization with photoelectron energies close to $2U_p$ was explained as a branch-cut contribution.

An interesting feature is the appearance of fan-shaped structures and richer interference patterns if the Coulomb phase is included, even if the trajectories are kept Coulomb free.  This type of structure is absent in the PAD computed with the standard SFA [Fig.~\ref{fig:padcoulfree}(d)], whose fringes have been described analytically in \cite{Maxwell2017}. This agrees with our previous work \cite{Maxwell2018}, in which the presence of the Coulomb potential is directly associated with this type of structure in analytic models. 

\begin{figure}
	\includegraphics[width=0.5\textwidth]{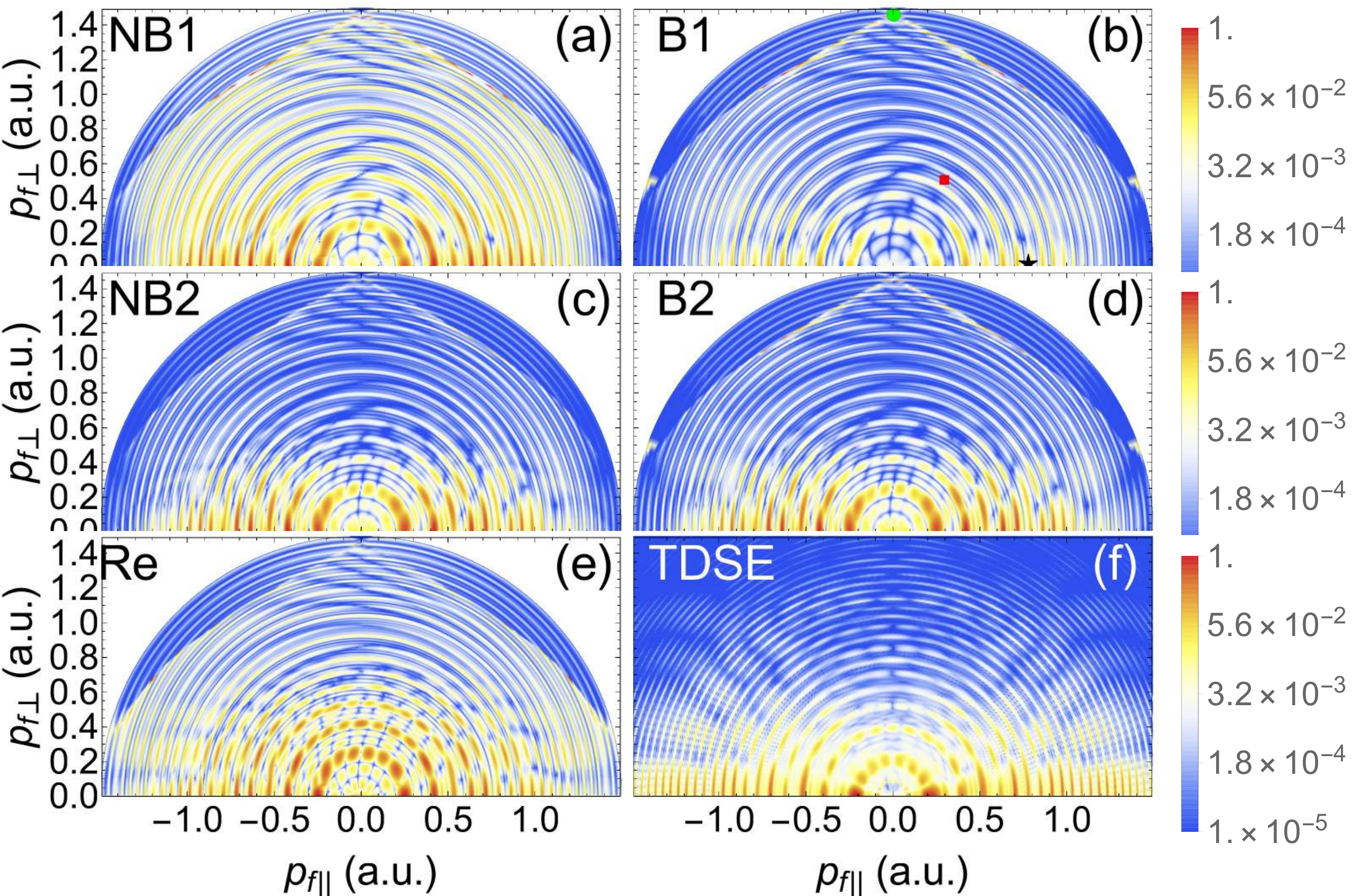}
	\caption{Photoelectron angular distributions computed using the CQSFA for Hydrogen ($I_p=0.5$ a.u.) in an external field of $\omega = 0.0570$ a.u.,Â Â Â  $U_p$ = 0.439Â Â Â  ($\lambda = 800 $nm, $I= 2 \times 10^{14} \mathrm{W}/\mathrm{cm}^2$), computed over four cycles. In panels (a) to (d), we present the outcome of the complex-trajectory CQSFA without [panels (a) and (c)] and with [panels (b) and (d)] branch cut corrections. Panel (e) displays the CQSFA result obtained with real trajectories, and panel (f) shows the outcome of an ab-initio computation, performed with the freely available software Qprop \cite{qprop,Mosert2016}.  In panels (c) and (d) the contributions of orbits 3 and 4 to the overall transition amplitude have been reduced multiplying by a factor 0.2. The red, green, and black dots corresponds to the momentum components  $(p_{f\parallel},p_{f\perp})=(-0.475, 0.400) $,   $(p_{f\parallel},p_{f\perp})=(-0.604,0.980) $  and $(p_{f\parallel},p_{f\perp})=(-0.619,0.0113)$ that have been used to compute the branch cuts in Fig.~\ref{fig:branchmapCQSFA}. All plots have been displayed in a logarithmic scale. 
	}
	\label{fig:PADCQSFA1}
\end{figure}

The distributions obtained in Fig.~\ref{fig:padcoulfree} are however very different from the full TDSE solution and do not reproduce the holographic patterns observed in experiments.  In Fig.~\ref{fig:PADCQSFA1}, we present the outcomes of different versions of the CQSFA, using complex trajectories without [Figs.~\ref{fig:PADCQSFA1}(a) and (c)] and with [Figs.~\ref{fig:PADCQSFA1}(b) and (d)] branch cut corrections, which are compared with the real-trajectory CQSFA and with the full TDSE solution [Figs.~\ref{fig:PADCQSFA1}(e) and (f), respectively]. 
All panels exhibit well-known interference structures such as the fan, the spider and the inter-cycle ATI rings. 
One should note, however, that the CQSFA overestimates the contributions of orbits 3 and 4, as can be seen in panels (a) and (b). 
This is particularly extreme if complex trajectories are used, and leads to a worse agreement with the TDSE for low photoelectron energies near the ionization threshold.  
If, on the other hand, the contributions of these trajectories are artificially reduced [panels (c) and (d)], the agreement with the TDSE considerably improves. 
The slope of the spider-like structure, which stems from the interference of orbits 2 and 3 \cite{Maxwell2017}, approaches its TDSE counterpart if complex trajectories are used. In contrast, the spiderlike fringes obtained with real trajectories are nearly horizontal.  The figure also reveals a pronounced spiral-like structure near the $p_{\perp}$ axis, which are caused by type 4 orbits, and a caustic that marks a boundary for which the contributions from orbit 3 are valid. Details have been provided elsewhere \cite{Maxwell2018}. 

\begin{figure}
	\includegraphics[width=0.5\textwidth]{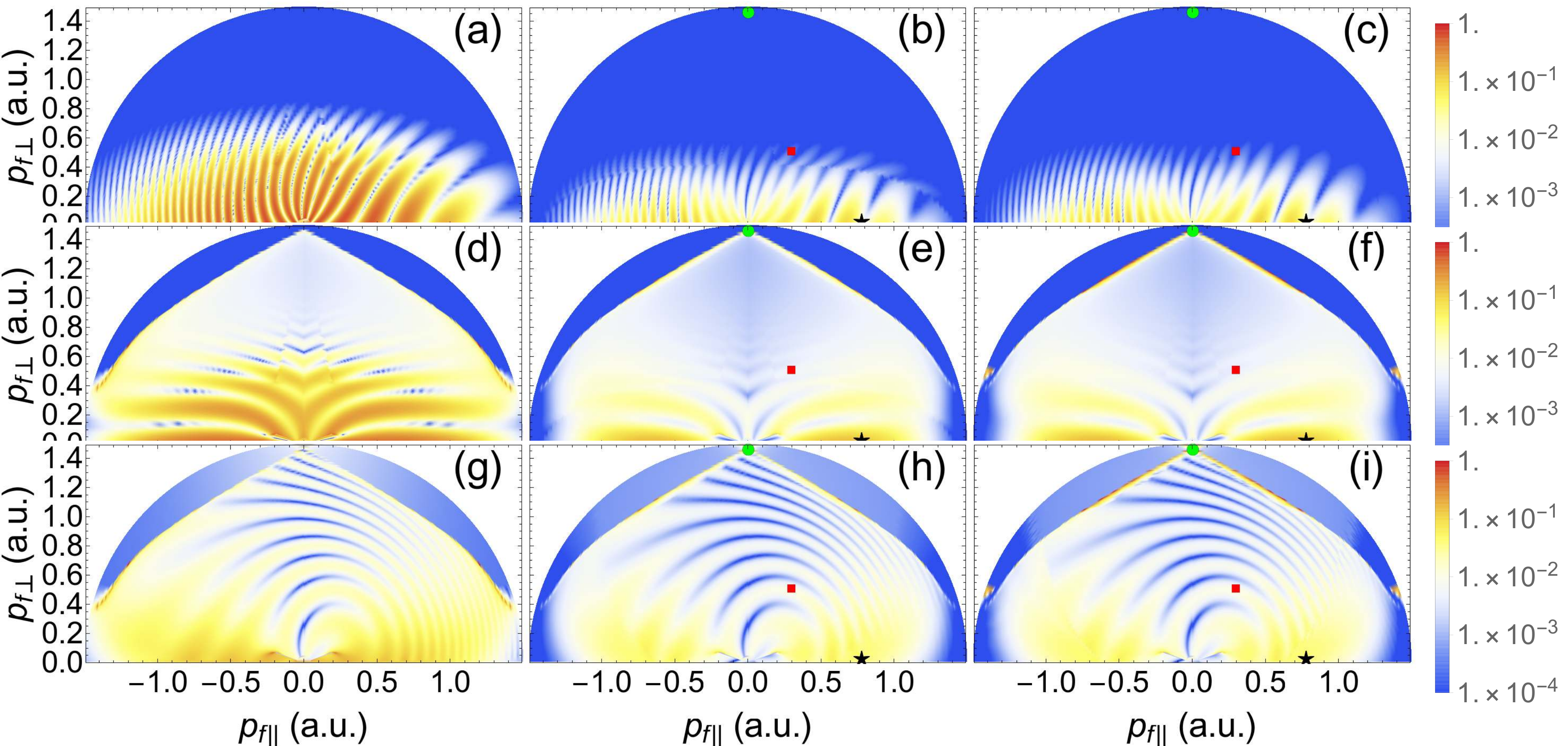}
	\caption{Contributions from specific pairs of orbits to the ATI PADS computed using different versions of the CQSFA for the same field and atomic parameters as in the previous figure. In the left, middle and right panels, we have employed real trajectories, complex trajectories and no branch cut corrections, and complex trajectories with branch-cut corrections, respectively. The first row [panels (a) to (c)] considers orbits 1 and 2, the middle row [panels (d) to (f)] orbits 2 and 3, and the lower row [panels (g) to (h)] to orbits 3 and 4.  The red, green, and black dots corresponds to the momenta $(p_{f\parallel},p_{f\perp})=(-0.475, 0.400) $,   $(p_{f\parallel},p_{f\perp})=(-0.604, 0.980) $  and $(p_{f\parallel},p_{f\perp})=(-0.619, 0.0113)$ that have been used to compute the branch cuts in Fig.~\ref{fig:branchmapCQSFA}.  All plots have been displayed in a logarithmic scale.  }
	\label{fig:PADCQSFA2}
\end{figure}

Further insight is given by analyzing the interference patterns formed by distinct pairs of orbits. 
The resulting distributions are depicted in Fig.~\ref{fig:PADCQSFA2}. 
The upper panels of the figure show the fan-shaped structure that results from the interference of orbits 1 and 2. 
For complex trajectories and when no contributions from the branch cuts are accounted for, the resulting fringes exhibit a discontinuous slope, which once more is related to crossing a branch cut [Fig.~\ref{fig:PADCQSFA2}(b)]. In fact, this discontinuity is eliminated if branch-cut corrections are incorporated [Fig.~\ref{fig:PADCQSFA2}(c)]. 
An interesting feature is that the PADs computed using complex trajectories decay much faster with increase of the photoelectron momentum perpendicular to the polarization direction than their real counterparts in this case. 
This is also a feature observed for the Coulomb-free trajectories [see Fig.~\ref{fig:Coulombfree}]. 
This faster decay is in agreement with what is observed for the TDSE computation [see Fig.~\ref{fig:PADCQSFA1}(d) for comparison]. 

The middle panels of the figure show the spider-like structure stemming from the interference of orbits 2 and 3. For real trajectories, the contrast of the spider-like fringes is much higher, while complex trajectories introduce some blurring.  This is caused by the fact that the imaginary parts employed in this model strengthen the contributions of orbit 3 and suppress those of orbit 2.  
The discontinuities near the $p_{\perp}$ axis, which are due to the approximations introduced in the sub-barrier corrections, also become smoother for complex trajectories. 
This is due to the orbit 2 contributions which decay faster with increasing $p_{\perp}$. Moreover, a direct comparison of Fig.~\ref{fig:PADCQSFA2}(d) with Figs.~\ref{fig:PADCQSFA2}(e) and (f) shows that the imaginary parts of the trajectories introduce a slope in the spider-like fringes. In fact, if real orbits are taken, such fringes are nearly horizontal. This difference in slopes has been identified in previous publications \cite{Maxwell2017,Maxwell2018}, in comparisons with ab-initio computations, but in that case the explanation remained speculative.  
Finally, complex trajectories also influence the spiral-shaped fringes that result from the interference of orbits 3 and 4, by altering their contrast and spacing. One should note that some of the corrections are in the vicinity of caustics and thus are obfuscated by them. 

\section{Conclusions}
The main conclusion to be drawn from this work is that, in quantum-trajectory approaches that take the residual Coulomb potential into consideration, the artifacts caused by branch cuts that occur in the ATI transition amplitude due to the specific functional form of the Coulomb potential can be corrected without prior mapping. 
In fact, it suffices to pick out a particular contour, test for branch cuts and distort this contour only if a branch cut is found. 
For the specific contour employed in this article, it was necessary to identify the intersection of the cuts with the real time axis.
After that the branching point can be identified employing a simple one-dimensional procedure of traveling along the cut side with a simultaneous calculation of the branch cut contribution into the complex-valued Coulomb phase. 
The present procedure is computationally much less demanding than existing methods, in which  all branch cuts are mapped in the complex time plane and a contour that avoids them is constructed subsequently \cite{Pisanty2016}.  
Furthermore, it also allows for complex Coulomb-distorted trajectories, instead of taking them to be real. 
The latter simplification is widely used, but is not consistent and renders the transition amplitudes dependent on the integration contour. 

The Coulomb potential is incorporated as an additional phase in the semiclassical action, and there are two options as far as the electron trajectoires are concerned. 
Either they are kept Coulomb-free and determined by the strong-field approximation, or the Coulomb force is incorporated and the full equations of motion of the electron in the continuum are solved. 
Physically, Coulomb-free trajectories are a reasonable approximation for circularly polarized driving fields, or driving fields of high ellipticity.  In contrast, for linearly polarized driving fields the Coulomb potential significantly modifies the trajectories and the distinction between direct and rescattered ATI is blurred \cite{Maxwell2018}. 

We tested our method by calculating ATI photoelectron angular distributions (PADs) using both types of trajectories. 
In both Coulomb-free and Coulomb-distorted cases, branch cuts lead to discontinuities in the fringes of holographic structures, which are corrected when our procedure is implemented. For the Coulomb-free case, our method provides a consistent framework, except for the limit of $p_{\bot}=0$ when the branch cut contribution into the Coulomb integral becomes divergent and requires regularization.
Instead, for the Coulomb-distorted case additional approximations were required. 
This was due to the fact that solving the full complex equations of motion for the electron is a highly nontrivial problem. 
Our major assumption was to equate the imaginary parts of the times and of the coordinate $\mathbf{r}(\tau)$ to those of their Coulomb-free counterparts. 
In particular when integrating along the real time axis, this will lead to a constant imaginary part for $\mathbf{r}$, which is consistent with a real (final) momentum at the detector. 
For orbit 1, we have verified that this assumption corresponds to the asymptotic limit $t \rightarrow \infty$ by solving numerically equations of motion in complex space and time.

Overall, complex trajectories lead to a faster decay with regard to increasing momentum components perpendicular to the field polarization axis, as compared to their real counterparts. This holds for both Coulomb-free and Coulomb-distorted trajectories, and supports the assertion in \cite{Torlina2013} that their imaginary parts cause a deceleration in the electronic wave packet. This suggests that the main influence of the imaginary parts is to change the weighting of the orbits and to change the overall yield. It seems that the main influence on the phases and interference patterns stems from their real parts. This may explain the success of the Coulomb-corrected methods in which such trajectories were taken to be real.

Comparisons with ab-initio methods also show that complex orbits improve the slope of the spider-like structure. Nonetheless, the overall agreement between the TDSE and the CQSFA worsens. This is due to the fact that the approximate imaginary parts of the coordinates $\mathbf{r}(\tau)$ introduced by our prescription overestimate the contributions of orbits 3 and 4. Throughout, approximating the Coulomb-distorted imaginary parts by their Coulomb-free counterparts seem to work better for type 1 and 2 orbits than for orbits 3 and 4. Physically, this is not surprising as the two former types of orbits are much closer to the SFA orbits obtained for direct electrons, while for the latter two deflection, and even rescattering plays a much more important role \cite{Maxwell2018,Maxwell2017a}. 

It is noteworthy that, within the present framework, a branching point is always associated with a return to the core, and, in the case of Coulomb-distorted trajectories, deflection or rescattering. 
In terms of rescattering in real time and space, gaps between the pairs of branching points can be associated with a nonzero value of the impact parameter for the photoelectron experiencing scattering.
Mathematically it makes however no difference if an integration contour circumvents a branching point tightly as in the algorithm presented here or it crosses  the gap in between the two branching points at its middle point as in Ref.~\cite{Pisanty2016}.
If both return and rescattering occurs at an nonvanishing angle, the effective impact parameter appears essentially nonzero, and the present method is applicable. 
There are however problems for final momenta near the polarization axis. 
Physically, this specific scenario would correspond to hard scattering, which has proven to be a challenge for the CQSFA already if real trajectories are used \cite{Maxwell2017,Maxwell2017a,Maxwell2018}. 

Within the present approach, the Coulomb integral appears divergent
in the case of hard recollision when ${\bf r}(t_k)=0$, which is
particularly problematic in the Coulomb-free case, for which momentum is
conserved. In this case, all problematic trajectories are concentrated along the polarization direction where the probability is known to have a maximum. 
In order to eliminate the divergency, a matching with the phase of a stationary atomic scattering wave function has to be performed, along a method similar to that applied for the matching at the saddle point $t=t_s$ \cite{Perelomov1967,Popruzhenko2014a}.
For the case of high harmonic generation this procedure has been realized in \cite{Popruzhenko2018b} employing the Coulomb-free trajectories.
Extension of this method to  ATI and  Coulomb-distorted trajectories remains a serious challenge.
Still, the present results may be viewed as a road map towards the full computation and characterization of complex trajectories and of overcoming branch cuts in photoelectron holography. 

\section*{Acknowledgements}
S.V.P. thanks J. M. Rost for useful discussions and University College London for its kind hospitality, and acknowledges financial support of the Russian Science Foundation, grant No. 18-12-00476. CF.M.F. and A.S.M. acknowledge support from the UK Engineering and Physical Sciences Research Council (EPSRC), grant No. EP/J019143/1.

\end{document}